\documentclass{article}
\usepackage{amssymb}
\usepackage{amsmath,bm}

\title{The Equivalence Principle and the Constants of Nature}
\author{Thibault Damour}
\date{\it Institut des Hautes Etudes Scientifiques, 35, route de Chartres, 91440 Bures-sur-Yvette, France}

\begin{document}

\maketitle

\begin{abstract}
We briefly review the various contexts within which one might address the issue of ``why'' the dimensionless constants of Nature have the particular values that they are observed to have. Both the general historical trend, in physics, of replacing a-priori-given, absolute structures by dynamical entities, and anthropic considerations, suggest that coupling ``constants'' have a dynamical nature. This hints at the existence of observable violations of the Equivalence Principle at some level, and motivates the need for improved tests of the Equivalence Principle.
\end{abstract}

\section{Introduction}

The currently known laws of physics contain about twenty independent dimensionless ``constants''. For instance, some of the most important ones, for determining the world around us, are:
\begin{equation}
\label{eq1}
\alpha_{\rm EM} = \frac{e^2}{\hbar c} \simeq \frac{1}{137.035 \, 999 \, 7} \, ,
\end{equation}
\begin{equation}
\label{eq2}
\frac{m_p}{m_e} \simeq 1836.152 \, 672 \, ,
\end{equation}
\begin{equation}
\label{eq3}
\frac{G \, m_e \, m_p}{\hbar c} \simeq 3.216 \times 10^{-42} \, .
\end{equation}

\medskip

An important question is: What determines the values of these constants? According to Leibniz, one of the basic principles of rational thinking is the {\it  Principle of Reason}: ``Nihil est sine ratione'' (``Nothing is without a reason''). What could be the ``reasons'' behind the very specific numbers quoted in Eqs~(\ref{eq1})-(\ref{eq3}) above? We do not have any firm answer to this question. The aim of this note is to recall the various contexts and scenarios within which this question might be addressed. The main conclusion of our discussion will be that it is important to perform improved tests of the Equivalence Principle because these tests are one of our few windows on the physics which is possibly at work for selecting the constants of Nature.

\section{Are the constants constant?}

Einstein's theory of General Relativity has deeply transformed one aspect of the general framework of physics. Before 1915, both the structure of spacetime and the laws of local matter interactions were supposed to be ``rigid'', i.e. given once for all, as absolute structures, independently of the material content of the world. General Relativity introduced the idea that the structure of spacetime might be ``soft'', i.e. influenced by its material content. On the
other hand, one of the basic principles of General Relativity, the Equivalence Principle (EP), postulates that the laws of local physics, and notably the values of all the dimensionless coupling constants, such as $\alpha_{\rm EM}$ or $m_p / m_e$, must be kept ``rigidly fixed''. General Relativity thereby introduces an {\it asymmetry} between a soft, dynamical spacetime structure and a rigid, non-dynamical set of coupling constants.

\smallskip

This asymmetry was questioned by Dirac \cite{Dirac:1937ti} and Jordan \cite{Jordan}. Dirac phenomenologically assumed that the small dimensionless coupling $G \, m_e \, m_p / \hbar c$ of Eq.~(\ref{eq3}) varied proportionally to the inverse of the age of the universe, while Jordan (reviving generalizations of General Relativity \`a la Kaluza-Klein) essentially assumed that both $\alpha_{\rm EM}$ and $G$ could become spacetime fields $\varphi (t,{\bm x})$. Actually, the first author to clearly realize that Jordan's original theory implied that the fine-structure constant $\alpha_{\rm EM}$ had become replaced by a field $\varphi (t,{\bm x})$ was Fierz \cite{Fierz}. Fierz then pointed out that astronomical data (line spectra of galaxies) were putting rather strong constraints on the spacetime variability of $\alpha_{\rm EM}$, and suggested to restrict the original, two-parameter class of Jordan's ``varying constant'' theories to the special one-parameter class where the fine-structure constant $\alpha_{\rm EM}$ remains constant, but where the gravitational coupling $G$ is allowed to become a spacetime field. [This EP-respecting one-parameter Jordan-Fierz theory coincides with the tensor-scalar theory later studied by Brans and Dicke.]

\section{Varying constants and Equivalence Principle violations}

The considerations of Jordan and Fierz on field-theory models of varying constants attracted the attention of Dicke. In particular, Dicke realized the important fact that any theory in which the local coupling constants are spatially dependent will entail some violation of the (weak) Equivalence Principle (EP), namely some non-universality in the free-fall acceleration of bodies embedded in an external gravitational field. Dicke's general argument \cite{Dicke} is that the mass $m_i$ of a body, which is made (in view of $mc^2 = E_{\rm tot} = \sum_{\alpha} \, E_{\alpha}$) of many contributions, related to various interaction energies (strong, weak, electromagnetic; to which we can now add the Higgs interactions, responsible for the ``rest masses'' of the quarks and the leptons), is a certain, complicated function of various coupling constants, notably the gauge and Yukawa coupling constants: $m_i = m_i [\alpha_{\rm EM} , \ldots]$. If the coupling constants are spatially dependent, the free-fall acceleration deduced from the action of a point particle embedded in a (general relativistic) gravitational field $g_{\mu\nu} (x)$,
\begin{equation}
\label{eq4}
S_{m_i} = - \int m_i [\alpha_{\rm EM} (x) , \ldots] \sqrt{-g_{\mu\nu} (x) \, dx^{\mu} \, dx^{\nu}} \, ,
\end{equation}
will read (in the slow-velocity limit)
\begin{eqnarray}
\label{eq5}
{\bm a}_i &= &{\bm g} - {\bm \nabla} \ln m_i [\alpha_{\rm EM} (x) , \ldots] \nonumber \\
&= &{\bm g} - \frac{\partial \ln m_i [\alpha_{\rm EM} , \ldots]}{\partial \, \alpha_{\rm EM}} \, {\bm \nabla} \, \alpha_{\rm EM} - \ldots \, .
\end{eqnarray}
The coefficients associated to the spatial gradients of the various coupling constants in Eq.~(\ref{eq5}) are expected not to be universal, so that ${\bm a}_i \ne {\bm a}_j$ if the composition of body $i$ differs from that of body $j$.

\smallskip

To turn the result (\ref{eq5}) into an explicit prediction for the composition dependence of the EP-violation parameter
\begin{equation}
\label{eq6}
\eta_{ij} \equiv \frac{a_i - a_j}{\langle a \rangle}
\end{equation}
one needs: (i) an explicit, Jordan-type, field model of the spacetime variability of coupling constants (predicting both the dynamics of the field $\varphi (t,{\bm x})$, and the dependence of $\alpha_{\rm EM}$, $m_p / m_e$, $G \, m_e \, m_p , \ldots$, on the field $\varphi$), and (ii) an estimate of the dependence of the mass of a body $i$ (say a chunk of Beryllium) on the various coupling constants of particle physics. Concerning the point (i), several models have been considered in the literature: the original (Kaluza-Klein-)Jordan-type scalar field, coupling (in the ``Einstein frame'') {\it only} to the electromagnetic action, and thereby affecting only $\alpha_{\rm EM}$, has been revived by Bekenstein \cite{Bekenstein:1982eu}. The properties of this model have been studied by several authors (e.g. \cite{Sandvik:2001rv}). Other authors have focussed on the more general type of field models suggested by String Theory, i.e. on ``dilaton models'' where a scalar field $\varphi (t,{\bm x})$ monitors, in a correlated manner, the spacetime variability of, essentially, all the coupling constants: gauge couplings, Yukawa couplings, gravitational coupling,$\ldots$. In these models, because of the complex dependence of mass on the various couplings (point (ii)), the EP-violation parameter (\ref{eq6}) has, in general, a complicated dependence on the nuclear composition of bodies $i$ and $j$ (see \cite{Damour:1994zq}). The dependence of the mass $m_i$ on quark masses, via nuclear interactions, is especially difficult to estimate, see \cite{Flambaum:2002wq,Donoghue:2006du,Damour:2007uv,Dent:2008gu}. The complexity of the composition-dependence of the EP-violation $\eta_{ij}$ in dilaton models is a phenomenologically interesting fact which might allow, in principle, to experimentally probe the existence of a long-range dilaton-like field, via EP tests comparing several different pairs of bodies. Correlatively, the predicted general structure of the composition dependence of $\eta_{ij}$ can be used to optimize the choice of materials in EP experiments \cite{DamourBlaser,Damour:1996xt}.

\smallskip

For reviews of the observational and theoretical status of ``varying constants'' see \cite{Uzan:2002vq,MartinsJENAM,Dent:2008us}, and the popular books \cite{BarrowBook}, \cite{FritzschBook}. See also the contributions of J.P.~Uzan and N.~Kolachevsky to these proceedings. Let us only mention the recent progress in the comparison of single-ion optical clocks which has allowed one to constrain the variability of $\alpha_{\rm EM}$ at the $10^{-17} \, yr^{-1}$ level \cite{Rosenband:2008} (thereby ``beating'' the indirect constraint on $d \ln \alpha_{\rm EM} / dt$ coming from the analysis of the Oklo phenomenon \cite{Shlyakhter:1976,Damour:1996zw,Fujii:1998kn}).

\section{Scenarios for the selection of coupling constants}

Let us come back to the main issue of concern: what could be the ``reason'' behind the specific values taken by the coupling constants of Nature, such as Eqs.~(\ref{eq1})-(\ref{eq3})? We have recalled above that, in General Relativity, the EP represents an obstacle towards finding such a ``reason''. Indeed, if the coupling constants are ``God-given'', absolute constants, there is no hope of finding, within Nature, a rational ground for selecting specific values for them. On the other hand, the history of physics suggests that there are no {\it absolute structures} in physics. Einstein taught us that the concepts of absolute space and absolute time were just approximations to a particular, cosmologically selected, {\it dynamical} spacetime $g_{\mu\nu} (x)$. Kaluza, Klein, Dirac and Jordan generalized the message of General Relativity by suggesting that both the fine-structure constant $\alpha_{\rm EM}$ and the gravitational constant $G$ might be, like $g_{\mu\nu} (x)$, {\it dynamical} entities, determined, or at least influenced, by the material content of the universe, and, in particular, by its cosmological evolution. Pauli realized (in his 1953 work which generalized the Kaluza-Klein mechanism to a higher-dimensional ``gauge symmetry group''; see \cite{Straumann:2008aa}) that $SU(n)$ gauge couplings $g_n^2$ could, like $\alpha_{\rm EM}$, be dynamical entities. Brout-Englert, Guralnik-Hagen-Kibble, and Higgs taught us that the masses of leptons and quarks could be dynamically determined by a (spontaneous symmetry-breaking) mechanism in which a certain auxiliary (``Higgs'') field $\phi (x)$ settles to the bottom of its potential well: $\phi (x) \to \phi_0$ (modulo some symmetry operation). String theory appears to be a vast generalization of the Einstein-Kaluza-Klein-Jordan-Pauli idea in which {\it all} the coupling parameters of the world around us (gauge couplings, Yukawa couplings, gravitational coupling,$\ldots$) are {\it dynamical} entities, related to the mean values (in the ``vacuum'' around us) of certain ``moduli fields'' $\varphi_A (x)$.

\smallskip

This historical trend suggests that all the numbers in Eqs.~(\ref{eq1})-(\ref{eq3}), and similar equations dealing with other couplings and ratios, might be determined by some dynamical mechanism in which some fields $\varphi_A (x)$ (which determine the local values of the coupling constants) acquire some, approximately constant and uniform, values in the ``vacuum'' around us. This opens up the hope of finding the ``reason'' why the numbers in Eqs.~(\ref{eq1})-(\ref{eq3}) have the values they take.

\smallskip

Initially, string theorists hoped that the stringent consistency requirements of string theories would somehow select a unique, stable ``vacuum'', in which consistency requirements and energy minimization would oblige the moduli fields $\varphi_A (x)$ determining the coupling constants of low-energy physics to take particular values $\langle \varphi_A (x) \rangle = \varphi_A^0$. This would be a striking vindication of Leibniz's Principle of Reason. So far it has not been possible to uncover such stringent vacuum-selecting consistency requirements. As a substitute to this grand hope of finding a {\it unique} consistent vacuum, many string theorists hope that there exists a {\it ``discretuum''} of consistent string vacua, i.e. a discrete set of vacua, in each of which the moduli fields take particular values $\varphi_A^0$, corresponding to some discrete, local minimum of the total energy (for recent reviews see \cite{Douglas:2006es,Denef:2008wq}). If that is the case, this would predict that the coupling constants do not have any temporal or spatial variability because, like in the Higgs mechanism, a fluctuation $\delta \varphi_A (x) = \varphi_A (x) - \varphi_A^0$ has an energy cost $\delta V (\varphi_A) \simeq \frac{1}{2} (\partial^2 V / \partial \varphi_A^0 \partial \varphi_B^0) \, \delta \varphi_A \, \delta \varphi_B$ which implies that $\delta \varphi_A (x)$ is a massive, short-ranged field (with Yukawa-type, exponentially suppressed effects). Though such a mechanism might entail observable short-range modifications of gravity \cite{Antoniadis:2007uz}, it predicts the absence of any long-range EP violations. Note that, far from providing no motivation for EP tests, the current majority view of string theorists does imply that EP tests are important: indeed, they represent tests of a widespread theoretical assumption, that any EP-violation observation would refute, thereby teaching us a lot about fundamental issues\footnote{I thank Mike Douglas for suggesting this positive way of formulating the potential theoretical impact of EP tests within the current string-theory majority view.}.

\smallskip

On the other hand, as the current attempts at stabilizing all the string-theory moduli fields (see, e.g., \cite{Denef:2008wq}) are extremely complex and look rather unnatural, one cannot help thinking that there might exist other ways in which string theory (or whatever theory reconciles General Relativity with Particle Physics) connects itself with the world as we observe it. In particular, we know that one of the (generalized) ``moduli fields'', namely the Einsteinian gravitational field $g_{\mu\nu} (x)$, plays a crucial role in determining the structure of the particle physics interactions via the fact that, in a local laboratory, one can approximate, to a high accuracy, a spacetime varying $g_{\mu\nu} (x)$ by a constant Poincar\'e-Minkowski metric $\eta_{\mu\nu}$. In other words, when listing the dimensionless coupling constants (\ref{eq1})-(\ref{eq3}),$\ldots$, of particle physics one should include $\eta_{\mu\nu} = {\rm diag} \, (-1 , +1 , +1 , +1)$ in the list, and remember that it comes from a long-range, cosmologically evolving field $g_{\mu\nu} (x)$. In this connection, let us further recall that the ``dilaton'', $\Phi (x)$, i.e. the moduli field which determines the value of the basic, ten-dimensional string coupling constant $g_s$ can be viewed (\`a la Kaluza-Klein) as an additional metric component $g_{11 \, 11} (x)$, measuring the size of a compactified eleventh dimension \cite{Witten:1995ex}. This family likeness between the dilaton $\Phi (x)$ and the metric $g_{\mu\nu} (x)$ (which entails a correlated likeness, say in heterotic string theory, between $g_{\mu\nu} (x)$ and the gauge couplings $g_a^2 (x)$, as well as the string-frame gravitational coupling $G(x)$) suggests that there might exist consistent string vacua where some of the moduli fields are not stabilized, but retain their long-range, spacetime-dependent character. As recalled above, such a situation would entail long-range violations of the EP. How come such violations have not yet been observed, given the exquisite accuracy of current tests of the universality of free fall (at the $10^{-13}$ level \cite{Schlamminger:2007ht}) and of current tests of the variability of coupling constants \cite{Rosenband:2008}? A possible mechanism for reconciling a long-range, spacetime varying dilaton (or, more generally, moduli) field $\Phi (x)$ with the strong current constraints on the time or space variability of coupling constants is the {\it cosmological attractor} mechanism \cite{Damour:1992kf,Damour:1994zq,Damour:2002mi} (for other attempts at using cosmological dynamics to stabilize the moduli fields see \cite{Greene:2007sa} and references therein). A simple realization of this mechanism is obtained by assuming that all the coupling functions $B_A (\Phi)$ of $\Phi$ to the fields describing the sub-Planckian particle physics (inflaton, gauge fields, Higgs field, leptons, quarks,$\ldots$) admit a limit as $\Phi \to +\infty$ (``infinite bare strong coupling'') \cite{Veneziano:2001ah}. Under this very general, technically simple (but physically highly non trivial) assumption, one finds that the inflationary stage of cosmological expansion has the effect of naturally driving $\Phi$ towards values so large that the present observational deviations from General Relativity are compatible with all the current tests of Einstein's theory \cite{Damour:2002mi,Damour:2002nv}. This ``runaway dilaton'' mechanism also yields an interesting connection between the deviations from General Relativity and the amplitude of large-scale cosmological density fluctuations coming out of inflation. In particular, the level of EP violation is predicted to be
\begin{equation}
\label{eq7}
\eta \equiv \frac{\Delta a}{a} \sim 5 \times 10^{-4} \, k \left( \frac{\delta \rho}{\rho} \right)^{\frac{8}{n+2}} \, ,
\end{equation}
where $k = (b_F / (c \, b_{\lambda}))^2$ is a combination of unknown dimensionless parameters expected to be of order unity, and where $\delta \rho / \rho$ denotes the amplitude of large-scale cosmological density fluctuations, while $n$ denotes the exponent of the inflationary potential $V(\chi) \propto \chi^n$. Inserting the value observed in our universe, $\delta \rho / \rho \sim 5 \times 10^{-5}$, and the value $n=2$ corresponding to the simplest chaotic inflationary potential $(V(\chi) = \frac{1}{2} \, m_{\chi}^2 \, \chi^2)$, the rough prediction (\ref{eq7}) yields $\eta \sim k \times 10^{-12}$ which, given that $k$ is only constrained ``to be of order unity'', is compatible with current EP tests. Note that this runaway dilaton mechanism then predicts (if $n=2$) that a modest increase in the accuracy of EP tests might detect a non zero violation. Note also the rationally pleasing aspect (reminiscent of Dirac's large number hypothesis \cite{Dirac:1937ti}) of Eq.~(\ref{eq7}) which connects the level of variability of the coupling constants to cosmological features (see \cite{Damour:2002nv} for further discussion of this aspect), thereby explaining ``why'' it is so small without invoking the presence of unnaturally small dimensionless numbers in the fundamental Lagrangian.

\smallskip

The ``runaway dilaton'' mechanism just mentionned was formulated as a possible way of reconciling, within a string-inspired phenomenological framework, a ``cosmologically running'' massless\footnote{Let us note in passing that an interesting generalization of the cosmological attractor mechanism is obtained by combining the attraction due to the coupling of $\Phi$ (via $B_A (\Phi)$) to the matter density, to the effect of a quintessence-like potential $V(\Phi)$ \cite{Khoury:2003rn}.} dilaton with observational tests of General Relativity. Let us note that some authors \cite{Wetterich:2008sx,Rabinovici:2007hz} have suggested that the puzzle of having an extremely small vacuum energy $\rho_{\rm vac} \lesssim 10^{-123} (m_{\rm Planck})^4$ might be solved by a mechanism of spontaneous breaking of scale invariance of some (unknown) underlying scale-invariant theory. Under the assumption that scale-invariance is re-established only when a certain ``dilaton field''\footnote{Beware that, here, the name ``dilaton field'' refers to a field, say $\varphi$, connected to scale invariance. Such a field $\varphi$ is, a priori, quite different from the ``dilaton field'' $\Phi$ of string theory. Indeed, string theory, as we currently know it, contains a basic mass (and length) scale, even in the limits $\Phi \to - \infty$ ($m_s^{(D=10)} = 1 / \sqrt{\alpha'}$) or $\Phi \to + \infty$ ($m_{\rm Planck}^{(D=11)} = 1 / \ell_{\rm Planck}^{(D=11)}$).} $\varphi \sim \ln \chi \to \infty$, it seems \cite{Wetterich:2008sx} that a ``$\varphi$-dilaton runaway'' behaviour (technically similar to the $\Phi$-dilaton runaway mentionned above) might take place and entail similar observational violations of the EP.

\section{Dynamics versus Anthropics}

We have mentionned above various visions of the ``reason'' behind the selection of the observed values (\ref{eq1})-(\ref{eq3}),$\ldots$ of the coupling constants. The intellectually most satisfactory one (given the historical pregnancy of the Principle of Reason \cite{Heidegger}) would be the discovery of subtle consistency requirements which would select an essentially unique physico-mathematical scheme describing the only possible physical laws. In this vision, all the dimensionless numbers of Eqs.~(\ref{eq1})-(\ref{eq3}),$\ldots$ would be uniquely determined. [Note that the discovery of asymptotic freedom and dimensional transmutation, see \cite{Gross:1973id,Politzer:1973fx}, has opened the way to a conceivable rational explanation of very small dimensionless numbers, such as Eq.~(\ref{eq3}) (which baffled Dirac): they might be exponentially related to smallish coupling constants, along the model $\Lambda_{\rm QCD} / \Lambda \sim \exp (-8\pi b / g^2)$ where $g^2$ is a  gauge coupling constant considered at the (high-energy, cut-off) scale $\Lambda$.]

\smallskip

In absence of precise clues for realizing this vision, we are left with two types of less satisfactory visions. In one, the extremely vast ``landscape of string vacua'' can dynamically channel the coupling constants towards a discretuum of possible ``locally special values''. This leaves, however, open the problem of finding the ``reason'' why our world has selected one particular set of such, energy-minimizing locally special values. In the other, all (or some of\footnote{Indeed, one can evidently mix the two different scenarios.}) the coupling constants are, like the metric of spacetime around us, dynamically determined by some global aspects of our universe. Both visions contain a partial dynamical ``reason'' behind the selection of the coupling constants, but both visions leave also a lot of room to contingency (or environmental influences). Many authors have suggested that a complementary ``reason'' behind the selection of the coupling constants that we observe, might be the (weak) ``Anthropic Principle'', i.e. the tautological requirement that the physical laws and conditions around us must be compatible with the existence of information processing organisms able to wonder ``why'' the world around them is as it is. In other words, this is essentially an issue of Bayesian statistics: one should consider only a posteriori questions, rather than a priori ones. Though the appeal to such an a posteriori consistency requirement is intellectually less thrilling than the demand of a stringent a priori consistency requirement, it might have satisfied Leibniz. Indeed, Leibniz was one of the enthusiastic historical proponents of the {\it ``Principle of Plenitude''} \cite{Lovejoy} which considers that all logically possible ``things'' (be they objects, beings or, even, worlds) have a tendency to (and therefore {\it must}, if one does not want contingency -- be it God's whim --  to reign) {\it exist}. In addition, in spite of its tautological character, the anthropic consistency requirement does lead to some well-defined, and scientifically interesting (as well as challenging) questions. Indeed, the general scientific question it raises is: what would change in the world around us if the values of the coupling constants (\ref{eq1})-(\ref{eq3}), etc., would be different? In its generality, this is a very difficult question to address. Let us mention here some scientifically interesting partial answers. [For the fascinating issue of what happens when one varies the vacuum energy density (or cosmological constant) see Refs.~\cite{Weinberg:1987dv,Vilenkin:1994ua}, which predicted that one should observe a non zero $\rho_{\rm vac}$ {\it before} any data had solidly suggested it.] The ``Atomic Principle'' refers to the scientific study of the range of coupling parameters compatible with the existence of the periodic table of atoms, as we know it. In particular, one might ask what happens when one changes the ratio $m_q / \Lambda_{\rm QCD}$ of light quark masses (or the Higgs vacuum expectation value which monitors the quark masses)
to the QCD energy scale. This issue has been particularly studied by Donoghue and collaborators \cite{Agrawal:1997gf,Agrawal:1998xa,Donoghue:2007zz,Donoghue:2009me}. Recent progress \cite{Damour:2007uv} has shown that the existence of heavy atoms is quite sensitive to the light quark masses. If one were to increase the mass ratio $(m_u + m_d) / \Lambda_{\rm QCD}$ by about 40\%, all heavy nuclei would unbind, and the world would not contain any non trivial chemistry.

\smallskip

Coming back to the issue of EP violation, one might use the idea of a (partially) anthropic selection of coupling constants to predict that the Equivalence Principle should be violated at some level. Indeed, as in the case of the vacuum energy mentionned above, the observed values of the coupling constants (as well as that of their temporal and/or spatial gradients or variability) should only be required to fall within some life-compatible range, and one should not expect that they take any special, more constrained value, except if this is anthropically necessary. When one thinks about it, one can see some reasons why too strong a violation of the universality of free fall might drastically change the world as we know it, but, at the same time, one cannot see any reason why the EP should be rigorously satisfied. Therefore, one should expect to observe
\begin{equation}
\label{eq8}
\frac{\Delta a}{a} \sim \eta_* \ne 0 \, ,
\end{equation}
where $\eta_*$ is the maximum value of $\eta \equiv \Delta a / a$ tolerable for life \cite{DamourDonoghue2009}. It is a challenge to give a precise estimate of $\eta_*$, but the prediction (\ref{eq8}) gives an additional motivation for EP tests.

\section{Conclusions}

Despite its name, the ``Equivalence Principle'' (EP) is not one of the basic principles of physics. There is nothing taboo about having an observable violation of the EP. On the contrary, one can argue (notably on the basis of the central message of Einstein's theory of General Relativity) that the historical tendency of physics is to discard any, a priori given, absolute structure. The EP gives to the set of coupling constants (such as $\alpha_{\rm EM} \simeq 1/137.0359997$) the status of such an, a priori given, absolute structure. It is to be expected that this absolute, rigid nature of the coupling constants is only an approximation. Many theoretical extensions of General Relativity (from Kaluza-Klein to String Theory) suggest observable EP violations in the sense that the set of coupling constants become related to spacetime varying fields.

\smallskip

However, there is no firm prediction for the observable level of EP violation. Actually, the current majority view about the ``moduli stabilization'' issue in String Theory is to assume that, in each string vacuum, the coupling constants are fixed by an energy-minimizing mechanism which is generically expected to forbid any long-range violation of the EP. This, however, makes EP tests quite important: indeed, they represent crucial tests of a widespread key assumption of string-theory model building. This exemplifies how EP tests are intimately connected with some of the basic aspects of modern attempts at unifying gravity with particle physics.

\smallskip

Some phenomenological models (inspired by string-theory structures, or attempting to understand the cosmological-constant issue) give examples where the observable EP violations would (without fine-tuning parameters) be just below the currently tested level. Such (runaway dilaton) models comprise many different, correlated modifications of Einsteinian gravity ($\Delta a/a \ne 0$, $\dot\alpha_{\rm EM} \ne 0$, $\gamma_{\rm PPN} - 1 \ne 0 , \ldots$), but EP tests stand out as our deepest possible probe of new physics. Anthropic arguments also suggest that the EP is likely to be violated at some (life-tolerable) level. Let us hope that the refined EP tests which are in preparation (see the contributions of F.~Everitt, P.~Touboul and M.~Kasevich to these proceedings) will open a window on the mysterious physics behind the selection of the coupling constants observed in our world.

\end{document}